# Bam-readcount - rapid generation of basepair-resolution sequence metrics

Ajay Khanna[1,†], David E. Larson[2,3,†], Sridhar Nonavinkere Srivatsan[1], Matthew Mosior[1,4], Travis E. Abbott[2,5], Susanna Kiwala[2], Timothy J. Ley[1,6], Eric J. Duncavage[7], Matthew J. Walter[1,6], Jason R. Walker[2], Obi L. Griffith[1,2,6,8], Malachi Griffith[1,2,6,8], Christopher A. Miller[1,6,*]

Affiliations:
1. Division of Oncology, Department of Internal Medicine, Washington University School of Medicine, St. Louis, MO
2. McDonnell Genome Institute, Washington University School of Medicine, St. Louis, MO
3. Current Affiliation: Benson Hill, Inc. St. Louis, MO
4. Current Affiliation: Moffitt Cancer Center, Tampa, FL
5. Current Affiliation: Google, Inc. Mountain View, CA
6. Siteman Cancer Center, Washington University School of Medicine, St. Louis, MO
7. Department of Pathology, Washington University School of Medicine, St. Louis, MO
8. Department of Genetics, Washington University School of Medicine, St. Louis, MO
† These authors contributed equally
* Corresponding Author

## Summary:

Bam-readcount is a utility for generating low-level information about sequencing data at specific nucleotide positions. Originally designed to help filter genomic mutation calls, the metrics it outputs are useful as input for variant detection tools and for resolving ambiguity between variant callers[1,2]. In addition, it has found broad applicability in diverse fields including tumor evolution, single-cell genomics, climate change ecology, and tracking community spread of SARS-CoV-2.[3–6].

## Availability and Implementation:

Here we report on the release of version 1.0 of this tool, which adds CRAM support, among other improvements. It is released under a permissive MIT license and available at https://github.com/genome/bam-readcount.

## Contact

c.a.miller@wustl.edu

**Introduction**

Though many tools exist that can call simple genotypes from sequence data, there is frequently a need for rapid and comprehensive reporting of sequencing metrics at specific genomic locations. The bam-readcount tool reports 15 metrics chosen specifically because they are known to be associated with the quality of sequence reads and individual base calls. These include summarized mapping and base qualities, strandedness information, mismatch counts, and position within the reads. This information can be useful in a large number of contexts, with one frequent application being variant filtering and ensemble variant calling situations where consistent, tool-agnostic metrics are useful[2,7,8].

**Implementation and results**

The ongoing adoption of compressed data formats has necessitated additions to the code, and the version 1.0 release that we report on here utilizes an updated version of HTSlib to support rapid CRAM file access[9]. This has also improved performance, and bam-readcount can report on 100,000 randomly selected sites from a 30x whole-genome sequencing (WGS) BAM in around 5 minutes[10]. It's performance scales nearly linearly with the number of genomic sites queried and average sequencing depth (Figure 1). Querying the same 100,000 sites from a BAM with 300x WGS takes 48 minutes, roughly 10x as long.

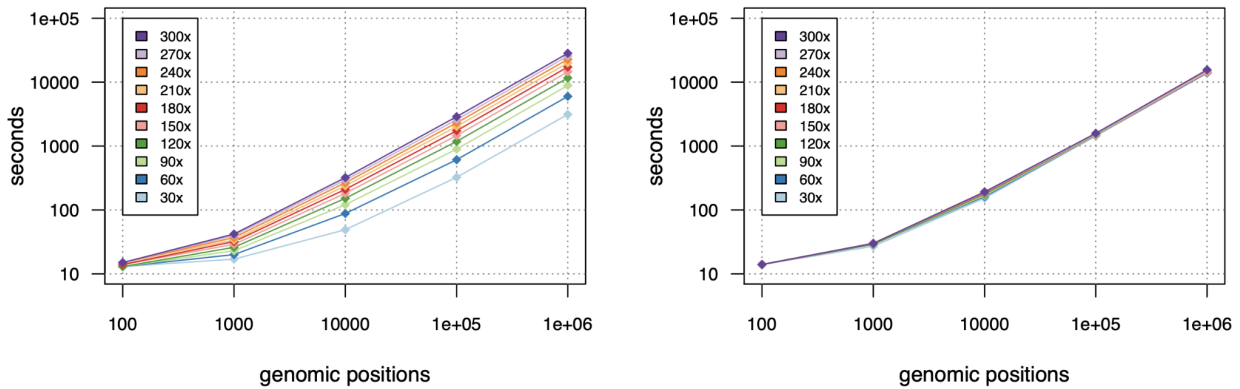

**Figure 1:** Performance of bam-readcount when querying randomly selected genomic positions from BAMs (left) or corresponding CRAMs (right) of varying sequencing depth. Colors correspond to average sequencing depth of the downsampled BAM/CRAM file.

Memory usage likewise is dependent on depth of sequencing, but still requires less than 1 GB of RAM for a 300x WGS BAM. Processing small CRAM files is somewhat slower than BAMs with comparable amounts of data, due to the increased CPU usage for decompression, but as depth increases, retrieval from disk becomes the bottleneck and operations on CRAMs exceed the speed of BAM. In our testing, on a fast SSD tier of networked disk, this transition occurs at a depth of about 180x. The problem is also embarrassingly parallel, so assuming adequate disk I/O, a roughly linear increase in speed can be achieved with a scatter/gather approach.

To lower barriers to adoption, we provide docker images for containerized workflows, and have developed a python wrapper that annotates a VCF file with read counts produced from this tool, (available as part of the VAtools package - https://github.com/griffithlab/VAtools).

**Conclusions**

Bam-readcount plays a central role in many genomic pipelines and there is a rich ecosystem of tools built on top of it that enable discovery. It has many uses in benchmarking and variant discovery, and it's feature-rich output has enabled deep learning approaches to variant calling and filtering[7,11]. In cancer genomics, it has been used for understanding pre-leukemic phenotypes and for detecting therapy-altering mutations from cell-free DNA[12,13]. Viral researchers have applied it to understand diversity in Varicella Zoster Virus Encephalitis and to perform epidemiological surveillance in wastewater of SARS-CoV-2[14,15]. Those with RNA-sequencing data have found it useful for identifying allele-specific expression in cancer, or for enabling copy-number detection in single-cell RNA sequencing[5,16]. It also serves as core infrastructure that supports genomics pipelines of all sizes, from bespoke workflows produced by small research groups to the NCI's Genomic Data Commons pipelines, where it has been run on tens of thousands of genomes[17–19].

Looking forward, we anticipate that as machine learning makes deeper inroads into genomics, the ability to extract highly informative features from large cohorts in a rapid manner will continue to make bam-readcount useful for the next generation of genomics research.

The bam-readcount tool is available at https://github.com/genome/bam-readcount and is shared under a MIT license to enable broad re-use.

**Data availability**

The WGS data used for benchmarking is available through dbGaP study phs000159, under sample id 452198/AML31. An archived snapshot of this code is available at ***zenodo***

**Authors' contributions**

Software Development: AK, DEL, SNS, MM, TEA, SK, CAM. Validation: AK, SNS, MM, CAM. Visualization: CAM. Supervision: CAM, MG, OLG, TJL, JRW, MJW  Funding Acquisition: CAM, MG, OLG, TJL, EJD, MJW.  Writing, review, and editing:  AK, DEL, SNS, MM, TEA, SK, TJL, EJD, MJW, JRW, OLG, MG, CAM

**Acknowledgements**

This work was supported by the National Cancer Institute [R50CA211782 to CAM, P01CA101937 to TJL, K22CA188163 to OLG, 1U01CA209936 to OLG, U24CA237719 to OLG], the Edward P. Evans Foundation (to MJW), and the National Human Genome Research Institute [R00 HG007940 to MG]